\documentclass[twocolumn,groupedaddress,prb]{revtex4}
\usepackage{amssymb}
\usepackage{bm}
\usepackage{amsmath}
\usepackage{graphicx}
\usepackage{color}

\begin{document}

\title{Isothermal-isobaric molecular dynamics using stochastic velocity rescaling}
\author{Giovanni Bussi}
\email{gbussi@unimore.it}
\altaffiliation{Present address: S3 Research Center
   and Dipartimento di Fisica, Universit\`a di Modena e Reggio Emilia, via Campi 213/A, 41100 Modena, Italy.
}
\author{Tatyana Zykova-Timan}
\author{Michele Parrinello}
\affiliation{Computational Science, Department of Chemistry and Applied Biosciences,
ETH Z\"urich, USI Campus, Via Giuseppe Buffi 13, CH-6900 Lugano, Switzerland}
\date{\today}

\begin{abstract}
 The authors present a new molecular dynamics algorithm for
 sampling the isothermal-isobaric ensemble.
 In this approach the velocities of all particles and volume degrees
 of freedom are rescaled 
 by a properly chosen random factor.
 The technical aspects concerning the derivation of the integration scheme and the
 conservation laws are discussed in detail.
 The efficiency of the barostat is examined in Lennard-Jones solid and liquid near
 the triple point and compared with the deterministic
 Nos\'{e}-Hoover and the stochastic Langevin methods.
 In particular, the dependence of the sampling efficiency on the choice
 of the thermostat and barostat relaxation
 times is systematically analyzed.
 \end{abstract}
\maketitle

\section{Introduction}

Starting from the breakthrough article of Andersen~\cite{ande80jcp}
many schemes have been proposed 
to modify the Newton equations of motion so as to perform molecular dynamics (MD)
in the isothermal-isobaric ensemble, where the number of particles, the
external pressure and the external temperature ($NPT$) are fixed.
Most of these schemes are based on an extended-Lagrangian formulation.
Before proceeding to overview the details of the extended system method,
it is better to separate two issues:
pressure and temperature control.
Within Andersen's framework the control of pressure was
achieved by adding to the Hamiltonian auxiliary dynamical variables,
which represented the volume $V$ and its conjugate momentum.
The resulting configurations were sampled from the $NPH$ ensemble, where
the enthalpy $H$ was constant.
This approach has been further extended to allow anisotropic variation
of the cell shape and size.\cite{parr+81jap,parr+82jcp}
Several almost-equivalent schemes have been proposed,
with similar performance and similar parameters,
often in combination with different thermostats
(see e.g.~Refs.~\onlinecite{hoov85pra,mart+94jcp}).

In the Andersen work, the temperature was regulated by additional
stochastic collisions, leading to a sampling of the $NPT$ ensemble.
The idea underlying the Andersen's thermostat was thus similar
to the use of Monte Carlo, and inherited from
Monte Carlo
the characteristic weak points such as discontinuity of the trajectories,
lack of a conserved quantity, at least in the original formulation, and
artifacts in the computation of dynamical properties.    
Another intrinsic disadvantage of the algorithm due to its stochastic nature
was the not-reproducibility of the trajectory, unless
the same random number generator was employed.

In subsequent refinements, the Nos\'{e}-Hoover
deterministic thermostat~\cite{nose84mp,nose84jcp,hoov85pra} 
replaced the stochastic one.
For this purpose, a new auxiliary variable was introduced, playing the role
of a friction on the particles and/or on the cell dynamics
and controlling the temperature.
This method allowed for the definition of a conserved quantity,
crucial to check the integration accuracy.
However, a weak point
was that the time-evolution of the kinetic energy was 
described by a second-order differential equation,
resulting in an algorithm which is inefficient for equilibration.
Even worst, the lack of ergodicity 
provided a poor control on the temperature in difficult
cases such as harmonic solids and, in general, systems with
normal modes.\cite{hoov86book}
Thermostat chains suggested in Ref.~\onlinecite{mart+92jcp} partially
solved these problems,
however at the price of a more complex formulation and a larger number of
parameters.

A possible alternative to the deterministic thermostat is stochastic
Langevin dynamics.\cite{schn-stol78prb} Its combination with extended-system control of the pressure,
the so-called Langevin piston scheme,\cite{langevin-piston} was proven to be efficient and ergodic
even in difficult cases.
The lack of a conserved quantity has been a traditional drawback
of stochastic MD, which was however recently solved.\cite{buss-parr07pre}
However, it is worth noting that the choice of the friction drastically influences the dynamics and, indirectly, the sampling efficiency.
On the one hand, a too small friction coefficient may cause a
very long equilibration time.
On the other hand, a too large friction coefficient may induce strong perturbations of the system dynamics hindering the exploration of the
phase space.

A further possibility for temperature and pressure control
is the deterministic scheme suggested by
Berendsen \emph{et al}.\cite{bere+84jcp}
This scheme is not based on an extended Lagrangian, and
the volume of the simulation box is driven by the difference between the internal pressure and
the target one.
Similarly, the kinetic energy is driven by its deviation from the
target one.
The algorithm is intuitive, easy to implement and efficient in the equilibration phase.
Unfortunately it cannot be used when
the correct distribution is required (e.g.~when fluctuations have to be calculated), 
since it does not sample the $NPT$ ensemble.
Moreover, in difficult cases the Berendsen scheme
systematically transfers energy to the slowest degrees of freedom and can lead to wrong
results.\cite{langevin-piston,harv+98jcc}
Hence the ``common wisdom'' suggests that the initial equilibration should be carried out with
the Berendsen scheme, while the ensemble averages should be calculated from the subsequent runs
performed with an extended-system method.

In a recent paper\cite{buss+07jcp} we proposed a thermostat which can be regarded as a stochastic version of the Berendsen
scheme.
Similarly to the Nos\'e-Hoover scheme,\cite{nose84mp,nose84jcp,hoov85pra}
it is a global thermostat, in the sense that it acts only
on the total kinetic energy of the system, and produces configurations in the
canonical ensemble.
In addition, this scheme allows defining a conserved quantity, does not suffer of ergodicity problems in solids~\cite{buss+07jcp} and
has been used successfully in practical applications
for equilibration purposes\cite{dona-gall07prl} or to perform
ensemble averages.\cite{brun+07jpcb,bard+08prl}
Finally, it has a minimal impact on the system dynamics, thus being an optimal choice for the computation
of dynamical properties.\cite{buss-parr08cpc}

The object of the present article is to combine the global stochastic thermostat
introduced in Ref.~\onlinecite{buss+07jcp}
with a standard barostat~\cite{ande80jcp,hoov85pra} to sample the isothermal-isobaric ensemble.
The properties of the resulting algorithm are discussed
in details and illustrated on Lennard-Jones (LJ) liquid and solid systems.
A special attention is dedicated to the parameterization of the thermostat and barostat relaxation times.
In particular, we provide a framework for the choice of the barostat relaxation time which
is based on the conservation of the total effective enthalpy,
thus on the accuracy of the resulting sampling.
For the thermostat relaxation time, we discuss its impact on the sampling efficiency by measuring
the autocorrelation times of several relevant physical quantities. In this sense, we
compare
the new scheme with the Nos\'{e}-Hoover\cite{nose84mp,nose84jcp,hoov85pra}
and with the Langevin piston barostats.\cite{langevin-piston}

This paper is organized as follows: In Section II we formulate
the most important aspects of the theoretical framework linking statistical theory
to equations of motion.
In particular, we demonstrate that the stochastic rescaling algorithm samples
the isothermic-isobaric distribution.
Section III contains several applications
of our scheme to LJ fluid and solid systems.
Some comments on the practical implementation of the barostat
and a reliable discretization scheme
are collected in the Appendix.

\section{Theory}\label{theory}

\subsection{Isothermal-isobaric ensemble}

We consider a system of $N$ particles,
described by coordinates $\bm{q}_i$, momenta $\bm{p}_i$, masses $m_i$,
contained in a box of variable volume $V$ 
subject to an external pressure $P_{ext}$.
With $q$ and $p$ we indicate the set of coordinates $\bm{q}_i$ and $\bm{p}_i$, respectively.
In the isotropic $NPT$ ensemble the probability distribution is usually defined as\cite{alle-tild87book,frenk-smit02book}
\begin{equation}
\label{eq-npt-probability}                            
\mathcal{P}_{NPT}(p,q,V)\propto e^{-\beta[K(p)+U(q,V)+P_{ext}V]},
\end{equation}
where $K(p)=\sum_i|\bm{p}_i|^2/(2m_i)$ is the kinetic energy, $U(q,V)$ is the potential energy
and $\beta=(k_BT)^{-1}$, with $k_B$ the Boltzmann constant and $T$ the temperature.
The expression for $\mathcal{P}_{NPT}$ has been addressed in the literature and is not unique.
For instance, Attard\cite{atta+95jcp}
proposed to use the logarithm of the volume as the integration variable,
or, equivalently, an extra $V^{-1}$ term in the probability distribution.
Nevertheless we will adopt the more standard definition in Eq.~\eqref{eq-npt-probability}.

When MD simulations are performed in absence of external forces
(i.e.~$\sum_j\partial U/\partial \bm{q}_j=0$)
the momentum of the center of mass of the system is conserved.
Thus, it is convenient to adopt
as dynamical variables the center-of-mass coordinate
$\bm{q}_{cm}$ and momentum $\bm{p}_{cm}$, and the
coordinates $\bm{r}$ and momenta $\bm{\pi}$ relative to the center of mass
\begin{subequations}
\label{eq-center-of-mass}
\begin{align}
\bm{q}_{cm}&=\frac{\sum_jm_j\bm{q}_j}{\sum_jm_j}
\\
\bm{p}_{cm}&=\sum_j\bm{p}_j
\\
\bm{r}_i&=\bm{q}_i-\bm{q}_{cm}
\\
\bm{\pi}_i&=\bm{p}_i-\frac{m_i\bm{p}_{cm}}{\sum_jm_j}.
\end{align}
\end{subequations}
After the change of variables, the $NPT$ ensemble reads
\begin{multline}
\label{eq-npt-probability1}                            
\mathcal{P}_{NPT}(\pi,r,p_{cm},q_{cm},V)\propto e^{-\beta[K(\pi)+K_{cm}(p_{cm})+U(r,V)+P_{ext}V]} \\
\times\delta\left(\sum_im_i\bm{r}_i\right)
\delta\left(\sum_i\bm{\pi}_i\right),
\end{multline}
where $K(\pi)=\sum_i|\bm{\pi}_i|^2/(2m_i)$ is the kinetic energy relative to the center of mass
and $K_{cm}(p_{cm})=|\bm{p}_{cm}|^2/(2\sum_im_i)$ is the kinetic energy of the center of mass.
The delta-functions in Eq.~\eqref{eq-npt-probability1} are related to the conservation of total momenta and inertia in
Eqs.~\eqref{eq-center-of-mass}
and to the fact that, in the center-of-mass framework,
$\sum_j m_j \bm{r}_j=0$
and $\sum_j\bm{\pi}_j=0$.
As a further simplification, the redundant variables $\bm{q}_{cm}$ and $\bm{p}_{cm}$
and can be integrated out, resulting in the distribution
\begin{multline}
\label{eq-npt-probability2}
\mathcal{P}_{NPT}(\pi,r,V)\propto Ve^{-\beta[K(\pi)+U(r,V)+P_{ext}V]} \\ \times
\delta\left(\sum_im_i\bm{r}_i\right)
\delta\left(\sum_i\bm{\pi}_i\right).
\end{multline}
Here the additional $V$ term comes from the fact that the size of the domain on which $\bm{q}_{cm}$
is integrated is proportional to $V$.
The contribution of the extra $V$ factor is almost negligible for systems with more than a few tens
of particles, though we will include it to have a formally correct ensemble.

The distribution in Eq.~\eqref{eq-npt-probability2} provides ensemble averages which are
equivalent to those from Eq.~\eqref{eq-npt-probability}, and should be used as a target
ensemble for any MD simulation in the $NPT$ ensemble when external forces are not present.
In the next two Subsections we provide an algorithm to sample this distribution.

\subsection{Pressure control}

We first consider the problem of fixing the pressure, adopting a solution similar
to that of Refs.~\onlinecite{ande80jcp,hoov85pra,mart+94jcp}.
Since we work in the reference frame of the center of mass, we choose an initial
configuration satisfying
$\sum_i\bm{\pi}_i=0$ and $\sum_im_i\bm{r}_i=0$.
Then we evolve it according to the following equations of motion
\begin{subequations}
\label{eq-nph-motion}
\begin{align}
\label{eq-nph-motion-r}
\dot{\bm r}_i=&\frac{\bm \pi_i}{m_i}+\eta {\bm r}_i
\\
\label{eq-nph-motion-p}
\dot{\bm \pi}_i=&{\bm f}_i-\eta {\bm \pi_i}
\\
\label{eq-nph-motion-eta}
\dot{\eta}=&\frac{3[V(P_{int}-P_{ext})+2\beta^{-1}]}{W}
\\
\label{eq-nph-motion-v}
\dot{V}=&3V\eta.
\end{align}
\end{subequations}
Here ${\bm f}_i=-\partial U/\partial {\bm r}_i$ are the forces,
$\eta$ is proportional to the relative change rate of the volume
and $W$ is the inertia of the piston, which determines the
typical time scale for volume changes.
The instantaneous internal pressure $P_{int}$
is given by
\begin{equation}
P_{int}=\frac{2K}{3V} - \frac{d U}{d V},
\end{equation}
where the derivative $d/dV$ is performed at fixed scaled coordinates, as discussed
in more detail in the Appendix.
Equations~\eqref{eq-nph-motion} are similar to those introduced by Hoover,\cite{hoov85pra}
but for an additional term in Eq.~\eqref{eq-nph-motion-eta}.
The original Hoover formulation, as observed by the author himself,
gives a slightly wrong ensemble.
With our correction,
when the dynamics is combined with a thermostat as shown in the next Subsection,
the exact $NPT$ distribution is sampled.
An alternative solution to this problem have been proposed by Martyna \emph{et al}.\cite{mart+94jcp}.

It can be easily shown that 
the following quantities are conserved during the dynamics:
\begin{subequations}
\label{eq-constant-of-motion-nph}
\begin{align}
\label{eq-constant-of-motion-nph-p}
\bm{\pi}_{cm}&=\sum_i\bm{\pi}_i=0 \\
\label{eq-constant-of-motion-nph-r}
\bm{r}_{cm}&=\frac{\sum_im_i\bm{r}_i}{\sum_im_i}=0 \\
\label{eq-constant-of-motion-nph-h}
H&=K(\pi)+U(r,V)-2\beta^{-1}\log V+P_{ext}V+\frac{W\eta^2}{2}=H_0.
\end{align}
\end{subequations}
The first two conservation laws come from the absence of external forces.
The last quantity $H$ is close to the enthalpy of the system ($K+U+P_{ext}V$),
and its initial value $H_0$ depends on the initial configuration and velocities.

When an ensemble of systems is evolved according to Eqs.~\eqref{eq-nph-motion},
their distribution evolves according to a corresponding
Liouville-like equation. Defining the $(6N+2)$-dimensional vector
$x=(r,\pi,V,\eta)$, Eqs.~\eqref{eq-nph-motion} can be written
in compact form as $\dot{x}=G(x)$, and the Liouville-like
equation on the density as $\dot{\rho}=-\partial_x(G \rho)=-\hat{L}\rho$,
where $\hat{L}$ is the Liouville operator.
Written explicitly in terms of the phase-space variables
$r$, $\pi$, $V$ and $\eta$, the Liouville operator reads
\begin{multline}
\label{eq-nph-liouville}
\hat{L}=
\sum_i\frac{\bm{\pi}_i}{m_i}\frac{\partial }{\partial \bm{r}_i}
+\sum_i\bm{f}_i\frac{\partial }{\partial \bm{\pi}_i}
+\eta \left(
\sum_i\bm{r}_i\frac{\partial }{\partial \bm{r}_i}
-  \sum_i\bm{\pi}_i\frac{\partial }{\partial \bm{\pi}_i}
\right)
\\
+\frac{3[V(P_{int}-P_{ext})+2\beta^{-1}]}{W}\frac{\partial}{\partial\eta}
+3\eta
+3V\frac{\partial}{\partial V}.
\end{multline}
The long time distribution for the ensemble can be found solving the equation
$\hat{L}\mathcal{P}=0$.
As it can be verified by direct substitution,
the following class of functions solves $\hat{L}\mathcal{P}_{NPH}=0$
\begin{multline}
\label{eq-nph-probability}
\mathcal{P}_{NPH}(\pi,r,V,\eta)\propto V^{-1}
\delta\left(\sum_im_i\bm{r}_i\right)
\delta\left(\sum_i\bm{\pi}_i\right) \\
\times\delta\left(
K+U-2\beta^{-1}\log V+P_{ext}V+\frac{W\eta^2}{2}-H_0
\right).
\end{multline}
The delta functions here come from the conservation laws in Eqs.~\eqref{eq-constant-of-motion-nph}.
Assuming that the system is ergodic an MD simulation based on
Eqs.~\eqref{eq-nph-motion} will produce configurations according to this distribution.
An alternative derivation of Eq.~\eqref{eq-nph-probability} can be done introducing the phase-space
compressibility as it is done for instance in Ref.~\onlinecite{tuck+01jcp}.

Rigorously speaking, Eq.~\eqref{eq-nph-probability} is not the standard $NPH$ distribution.
Indeed, here the constant of motion is $H$, which, as discussed above, slightly deviates from the enthalpy.
Moreover, there is an additional $V^{-1}$ prefactor. However, since we are interested in $NPT$ sampling,
we ignore these discrepancies and proceed in the combination of the $NPH$ scheme
with a thermostat.

\subsection{Temperature control}

We now combine the presented constant-enthalpy scheme with a thermostat.
We first extend the $NPT$ ensemble of Eq.~\eqref{eq-npt-probability2} so as to include $\eta$ as a dynamical variable.
Since $\eta$ is a velocity, we associate to it a kinetic energy and define the extended $NPT$ ensemble
as
\begin{multline}
\label{eq-npt-probability3}
\mathcal{P}_{NPT}(\pi,r,V,\eta)\propto V\delta\left(\frac{\sum_im_i\bm{r}_i}{\sum_im_i}\right) \\ \times
\delta\left(\sum_i\bm{\pi}_i\right) e^{-\beta[K^*(\pi,\eta)+U(r,V)+P_{ext}V]},
\end{multline}
where $K^*(\pi,\eta)=K(\pi)+\frac{W\eta^2}{2}$ is the total kinetic
energy, which includes also the barostat kinetic energy.
When the $\eta$ variable is integrated out, this distribution is
identical to the target one in Eq.~\eqref{eq-npt-probability2}.
Thus, we will use this distribution as a target ensemble for the $NPT$ algorithm.
It is easy to verify that this distribution is stationary with respect to the Liouville-like
operator in Eq.~\eqref{eq-nph-liouville}, i.e.~$\hat{L}
\mathcal{P}_{NPT}(\pi,r,V,\eta) = 0$.
The only reason why a time average performed with Eqs.~\eqref{eq-nph-motion}
will not give results consistent with the $NPT$ ensemble is that Eqs.~\eqref{eq-nph-motion}
have an extra constant of motion (namely $H$).
To sample the $NPT$ ensemble one has to introduce a thermostat,
which changes at every step
the value of $H$ in such a way that the $NPT$ ensemble is stationary.
Since $\mathcal{P}_{NPT}$ is the product of a term depending
on the potential energy and a term depending on kinetic energy, the thermostat
can be designed to act only on the latter one,
whose distribution in the Gibbs ensemble
is known \emph{a priori} and is
$\mathcal{P}_{NPT}(K^*)=(K^*)^{\frac{N_f}{2}-1} e^{-\beta K^*}$.
Here $N^*_f=3N-2$ is the number of degrees
of freedom, including the volume ($+1$) and excluding the center of mass ($-3$).
In the following, we describe how this task is accomplished by the
stochastic velocity rescaling.\cite{buss+07jcp}

In the stochastic velocity rescaling, the application of the thermostat
consists of a rescaling of the velocities of the system.
We here rescale not only the momenta $\pi$ but also the barostat velocity
$\eta$, by a factor $\alpha$:
\begin{subequations}
\begin{align}
{\bm \pi}_i\leftarrow&\alpha {\bm \pi}_i
\\
\eta\leftarrow&\alpha \eta.
\end{align}
\end{subequations}
This change affects the total kinetic energy $K^*=K+\frac{W\eta^2}{2}$, which
includes also the barostat kinetic energy.
Following Ref.~\onlinecite{buss+07jcp}, we calculate at each step the
rescaling factor by propagating
the total kinetic energy according to the equation
\begin{equation}
\label{eq-kinetic}
dK^*(t)=-\frac{K^*(t)-\bar{K}^*}{\tau_T}dt + 2\sqrt{\frac{\bar{K}^*K^*(t)}{N_f^*\tau_T}} dW(t).
\end{equation}
The propagation is done
for a half timestep $\Delta t/2$ as described in
the Appendix, so that
$\alpha^2=K^*(t+\Delta t/2)/K^*(t)$.
Here
$\tau_T$ is the relaxation time of the thermostat,
$\bar{K}^*=N^*_f/(2\beta)$ is the average kinetic energy,
and $dW$ is a Wiener noise.\cite{gard03book}
At variance with the scheme in Ref.~\onlinecite{buss+07jcp},
here the cell degree of freedom needs to
be counted, and the rescaling procedure affects both
the velocities of the particle and the velocity of the piston $\eta$.

As it has been shown previously,\cite{buss+07jcp} the stationary distribution associated
with the stochastic dynamics in Eq.~\eqref{eq-kinetic} is the canonical one,
i.e.~$(K^*)^{\frac{N_f}{2}-1} e^{-\beta K^*}$.
It immediately follows that, when the equation is combined with the constant-enthalpy dynamics,
one obtains configurations distributed according to the $NPT$ ensemble in Eq.~\eqref{eq-npt-probability3}.

Our algorithm consists of an alternation of constant-enthalpy steps [Eq.~\eqref{eq-nph-motion}] and 
a thermostat step [Eq.~\eqref{eq-kinetic}], as discussed in details in the Appendix.
It is worthwhile noting that other schemes can be implemented in
the same fashion,
simply by changing the thermostat step.
In particular, the Langevin piston algorithm\cite{langevin-piston} is easily implemented using the thermostat step
as discussed in Refs.~\onlinecite{adja+06epjb,buss-parr07pre}, applying the friction and noise to the particles and
the cell velocities, and removing the velocity of the center of mass afterwards so as
to eliminate the total force on the center of mass and obtain the same ensemble.
Similarly, a standard Nos\'{e}-Hoover scheme may be coupled with the constant-enthalpy integrator. 

\subsection{Control of sampling errors}\label{sec-sampling-errors}

In the practical implementation of MD one integrates
Eqs.~\eqref{eq-nph-motion} using a finite timestep.
This invariably introduces small systematic errors in the sampled
distribution.
The traditional way to control this error is to perform a microcanonical
simulation and to verify the conservation of the total energy.
This approach is easily extended to variable cell simulations
in the $NPH$ ensemble, where the conservation of the enthalpy is
used.
An equivalent conserved quantity can also be defined for simulations
performed with the Nos\'e-Hoover scheme in the $NPT$ ensemble.
In a recent paper we have introduced a formalism which allows
to define a conserved quantity, called effective enthalpy, also for stochastic dynamics.
This has been done for our stochastic velocity rescaling,\cite{buss+07jcp}
and has been later extended to Langevin dynamics.\cite{buss-parr07pre}
Here we discuss how this concept is generalized to an effective enthalpy in variable cell
simulations. To this aim, we first repeat some of the ideas introduced in
the previous papers,
then we show how the compressibility is taken
into account for our variable cell algorithm.

We consider the $(6N+2)$-dimensional vector $x=(p,r,\eta,V)$ in the phase space.
In principle MD equations would produce a continuous trajectory $x(t)$.
However, in the practical implementation the time is discretized
so that only a discrete sequence of snapshots of the system is
available, $x_0,x_1,\dots$, which is then used to evaluate ensemble averages.
In our scheme, each point $x_i$ is obtained from the previous one $x_{i-1}$ applying
alternatively Eq.~\eqref{eq-nph-motion} and the thermostat.
We now consider the move from $x_i$ to $x_{i+1}$ as a proposal for a Monte Carlo
move.
Even if in an MD simulation all the moves are accepted, it is instructive
to calculate the acceptance rate for that move.
To enforce detailed balance relative to the target distribution $\mathcal{P}_{NPT}(x)$,
the acceptance rate in the Metropolis scheme should be set equal to
\begin{equation}
\label{eq-detailed-balance}
\min
\left(
1,
\frac{M(x_{i}^*\leftarrow x_{i+1}^*)\mathcal{P}_{NPT}(x_{i+1}^*)}{M(x_{i+1}\leftarrow x_{i})\mathcal{P}_{NPT}(x_i)}
\right).
\end{equation}
The transition matrix $M(x'\leftarrow x)$ is the
probability of generating $x'$ as the next point in the sequence given that the
present point is $x$.
The star denotes time reversal of velocities, i.e.~if $x=(p,r,\eta,V)$
then $x^*=(-p,r,-\eta,V)$. Clearly, $\mathcal{P}_{NPT}(x)=\mathcal{P}_{NPT}(x^*)$.
The detailed balance enforced by Eq.~\eqref{eq-detailed-balance}
is not the standard one but its generalization for odd and even variables
discussed in Ref.~\onlinecite{gard03book}.
We then associate an
effective enthalpy $\tilde{H}_i$ to each point in the sequence
in such a way that the acceptance rate in Eq.~\eqref{eq-detailed-balance} is equal
to $\min
\left(
1,e^{-\beta(\tilde{H}_{i+1}-\tilde{H}_i)}
\right)$. By simple substitution, the change of effective enthalpy between subsequent points
is defined as
\begin{equation}
\tilde{H}_{i+1}-\tilde{H}_i
=
-\beta^{-1}
\ln
\frac{M(x_{i}^*\leftarrow x_{i+1}^*)\mathcal{P}_{NPT}(x_{i+1}^*)}{M(x_{i+1}\leftarrow x_{i})\mathcal{P}_{NPT}(x_i)}.
\end{equation}

We now proceed to an explicit evaluation of the increment of effective enthalpy at each
MD step.
Our scheme is a sequence of two different operations:
the thermostat and the time evolution of Eq.~\eqref{eq-nph-motion}.
When the thermostat is applied, either stochastic rescaling or Langevin,
its propagation is done analytically, so that detailed balance
is exactly satisfied. This gives an acceptance equal to 1 and
a vanishing change in the effective enthalpy.\cite{buss+07jcp}

On the other hand, when Eqs.~\eqref{eq-nph-motion} are propagated, the change in
effective enthalpy is not trivial and has to be evaluated explicitly.
The change in effective enthalpy is decomposed in two terms as
\begin{equation}
\label{eq-effective-energy-decomposed}
\tilde{H}_{i+1}-\tilde{H}_i
=
-\beta^{-1}
\ln
\frac{
\mathcal{P}_{NPT}(x_{i+1}^*)
}{
\mathcal{P}_{NPT}(x_i)
}
-
\beta^{-1}
\ln
\frac{
M(x_{i}^*\leftarrow x_{i+1}^*)
}{
M(x_{i+1}\leftarrow x_{i})
}.
\end{equation}
The first term, substituting the ensemble probability in Eq.~\eqref{eq-npt-probability3} 
and the definition of enthalpy in Eq.~\eqref{eq-constant-of-motion-nph-h}, is
\begin{multline}
\label{eq-effective-energy-probability}
-\beta^{-1}
\ln
\frac{
\mathcal{P}_{NPT}(x_{i+1}^*)
}{
\mathcal{P}_{NPT}(x_i)
}
=
H(x_{i+1})-H(x_i)
+ \\
\beta^{-1}
(\ln V_{i+1}-\ln V_i),
\end{multline}
where $V_i$ is the volume at step $i$.
The second term of Eq.~\eqref{eq-effective-energy-decomposed} is calculated
reminding that the rate between the backward transition matrix
and the forward transition matrix is equal to the Jacobian
of the transform which, as shown in the appendix, is
$V_{i+1}/V_i$. Thus
\begin{equation}
\label{eq-effective-energy-compression}
-
\beta^{-1}
\ln
\frac{
M(x_{i}^*\leftarrow x_{i+1}^*)
}{
M(x_{i+1}\leftarrow x_{i})
}
=
-\beta^{-1} (\ln V_{i+1}-\ln V_i).
\end{equation}
Combining Eqs.~\eqref{eq-effective-energy-decomposed}, \eqref{eq-effective-energy-probability}
and \eqref{eq-effective-energy-compression} one obtains:
\begin{equation}
\tilde{H}_{i+1}-\tilde{H}_i
=
H(x_{i+1})-H(x_i),
\end{equation}
so that the increment in effective enthalpy is exactly equal to the
increment in the enthalpy.

In conclusion, the calculation of the effective enthalpy is very similar
to the calculation of the effective energy in Ref.~\onlinecite{buss+07jcp}.
The effective enthalpy is obtained as a cumulated sum of all
the increments of $H$ in the constant-enthalpy dynamics.
Alternatively, one can obtain the effective enthalpy
as the total enthalpy $H$
minus the sum of all the contributions to the kinetic energy
given to the thermostat.
In the limit of zero time step, the effective enthalpy is exactly conserved.
In real applications,
the effective enthalpy is expected to exhibit random fluctuations and
a small systematic drift.
In principle, if exact sampling is required,
the change in effective enthalpy can be used to perform
hybrid Monte Carlo simulations,\cite{duan+97plb}
or to reweight properly the 
obtained configurations.\cite{wong-lian97pnas}
In practice, one can just monitor its systematic drift
and use it to quantify the sampling errors.
If the drift is too large, one should decrease the integration time step.

\section{Examples}\label{exam}
We now test the discussed algorithm on a couple of model systems.
We consider a system of 256 particles interacting through LJ
potential in a cubic box with periodic boundary conditions.
Simulations are carried out both in the solid fcc phase and in the liquid one
at temperature $\beta^{-1}=0.692$ and pressure $P_{ext}=0$, close to the triple point.
We use everywhere LJ reduced units.
The interactions are truncated at 2.5, and the force is smoothed
between 2.25 and 2.5.
Long-range corrections which take into account the difference
between the true LJ potential and our truncated approximation
are added to the total energy of the system.
To this aim, a constant radial-distribution-function is assumed
for distances larger than 2.25,\cite{alle-tild87book}
and the corrections to the internal pressure are evaluated as 
derivatives of the energy correction with respect to the cell volume.
The dynamics is computed using an in-house code.

For convenience, the piston mass $W$ is defined as
$N_f \beta^{-1}\tau_P^2$, where $\tau_P$ is a time describing 
the time-scale of the volume fluctuations.\cite{nose84jcp}
For the Nos{\'e}-Hoover scheme discussed in Section \ref{exam},
the thermostat mass is chosen
as $Q=N_f\beta^{-1}\tau_T^2$, where the thermostat 
relaxation time $\tau_T$ measures the thermostat strength.

In the following, three kinds of thermostats are considered:
(a) our scaling procedure (stochastic rescaling), which is stochastic
and global; (b) Langevin piston, which is stochastic and local;
(c) Nos{\'e}-Hoover, which is deterministic and global.
This choice allows us to investigate separately which is the
effect of a global thermostat versus a local one and
how the stochastic nature of the algorithm affects its performance.
We investigate the role played by the relevant simulation
parameters, namely the
timestep $\Delta t$,
thermostat relaxation time $\tau_T$,
the barostat relaxation time $\tau_P$.
For the Langevin scheme, we choose a friction
$\gamma=(2\tau_T)^{-1}$, for the reasons
discussed in Ref.~\onlinecite{buss-parr08cpc}.
The presented averages are obtained from $10^7$ steps simulations 
for each possible choice of the simulation parameters.

\subsection{Timestep, barostat mass and effective-enthalpy drift}

We consider here the choice of the timestep $\Delta t$ and of the barostat
relaxation time $\tau_P$.
As it will become clear later, the rationale behind the choice of these two parameters
is the same, thus we discuss them at the same time.

\begin{figure}[h]
\includegraphics[width=0.48\textwidth,clip]{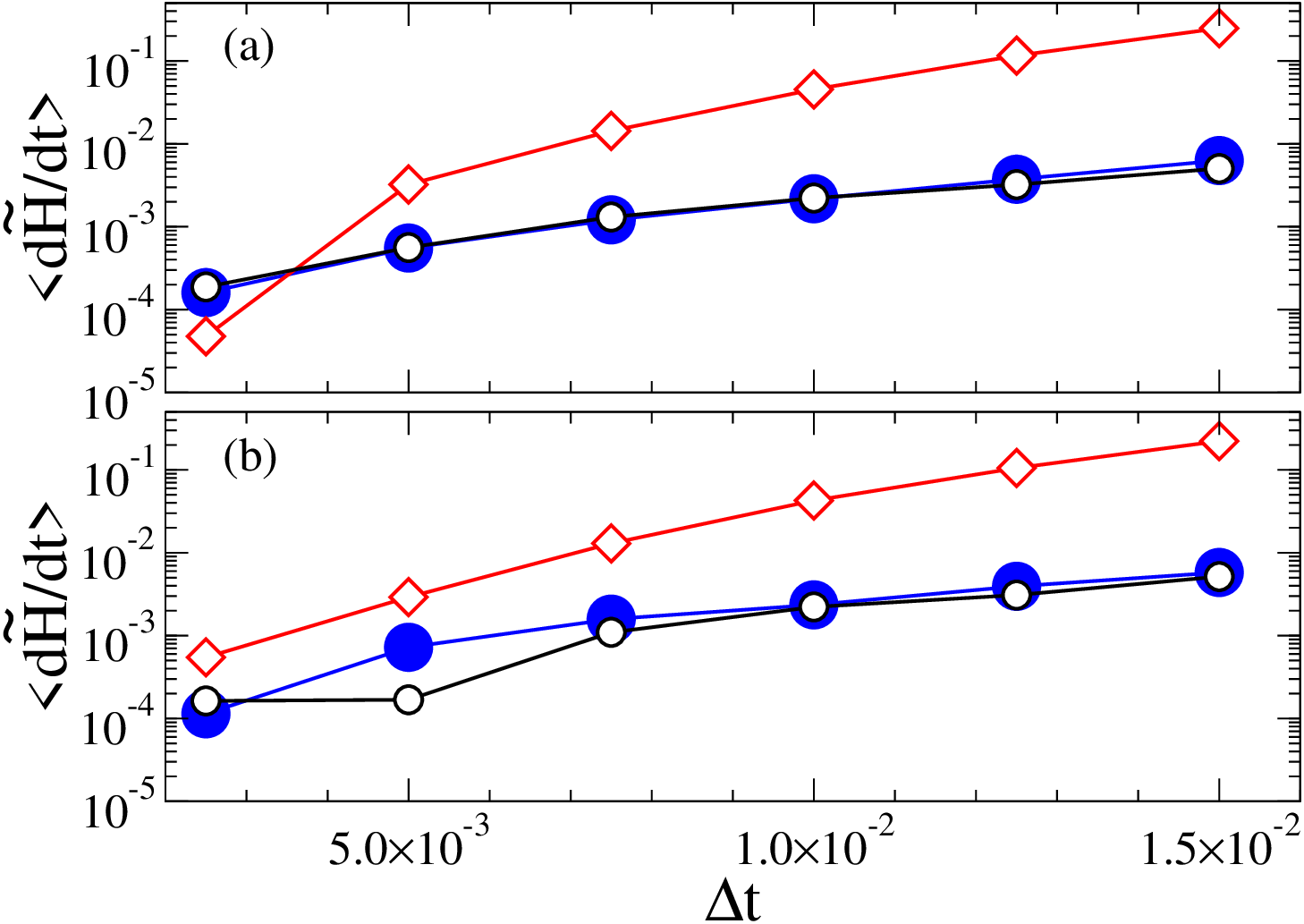}
\caption{Drift of the effective enthalpy $\langle\frac{d\tilde{H}}{dt}\rangle$ 
plotted as a function of the timestep $\Delta t$ for the Lennard-Jones liquid (a) and solid (b) systems at
the triple point, for stochastic velocity rescaling (filled circles),
Nos\'{e}-Hoover (open circles) and Langevin (diamonds) algorithms.
The thermostat relaxation time is $\tau_T=0.2$ and the barostat relaxation time
is $\tau_P=0.5$.
All quantities are in Lennard-Jones reduced units.}
\label{flux_dt}
\end{figure}

In Fig.~\ref{flux_dt} we show the dependence of the effective-enthalpy drift
$\langle \frac{d\tilde{H}}{dt}\rangle$ on $\Delta t$,
fixing the other parameters at $\tau_P=0.5$ and $\tau_T=0.2$.
The customary procedure in MD simulations is to choose
a $\Delta t$ which is as large as possible, with the constraint that the drift 
of the total energy is smaller than a given threshold.
In our case, an equivalent check can be done on the effective
enthalpy introduced in Subsection~\ref{sec-sampling-errors} or, for the Nos\'e-Hoover,
on its conserved quantity.
The exact value of the
threshold is rather arbitrary, but still it gives a quantitative estimate of the
sampling errors given by the finite timestep,
and still can be used to compare
the relative accuracy of two different simulations.
The two global schemes (Nos\'{e}-Hoover and stochastic rescaling) exhibit a similar drift, which
grows with $\Delta t$. Also for the Langevin scheme the drift increases with the
timestep, but faster than in the global schemes.
For further calculations, we choose a standard value $\Delta t=0.005$,
even if a slightly smaller timestep should be used
in order to achieve the same accuracy with Langevin dynamics.

\begin{figure}[h]
\includegraphics[width=0.48\textwidth,clip]{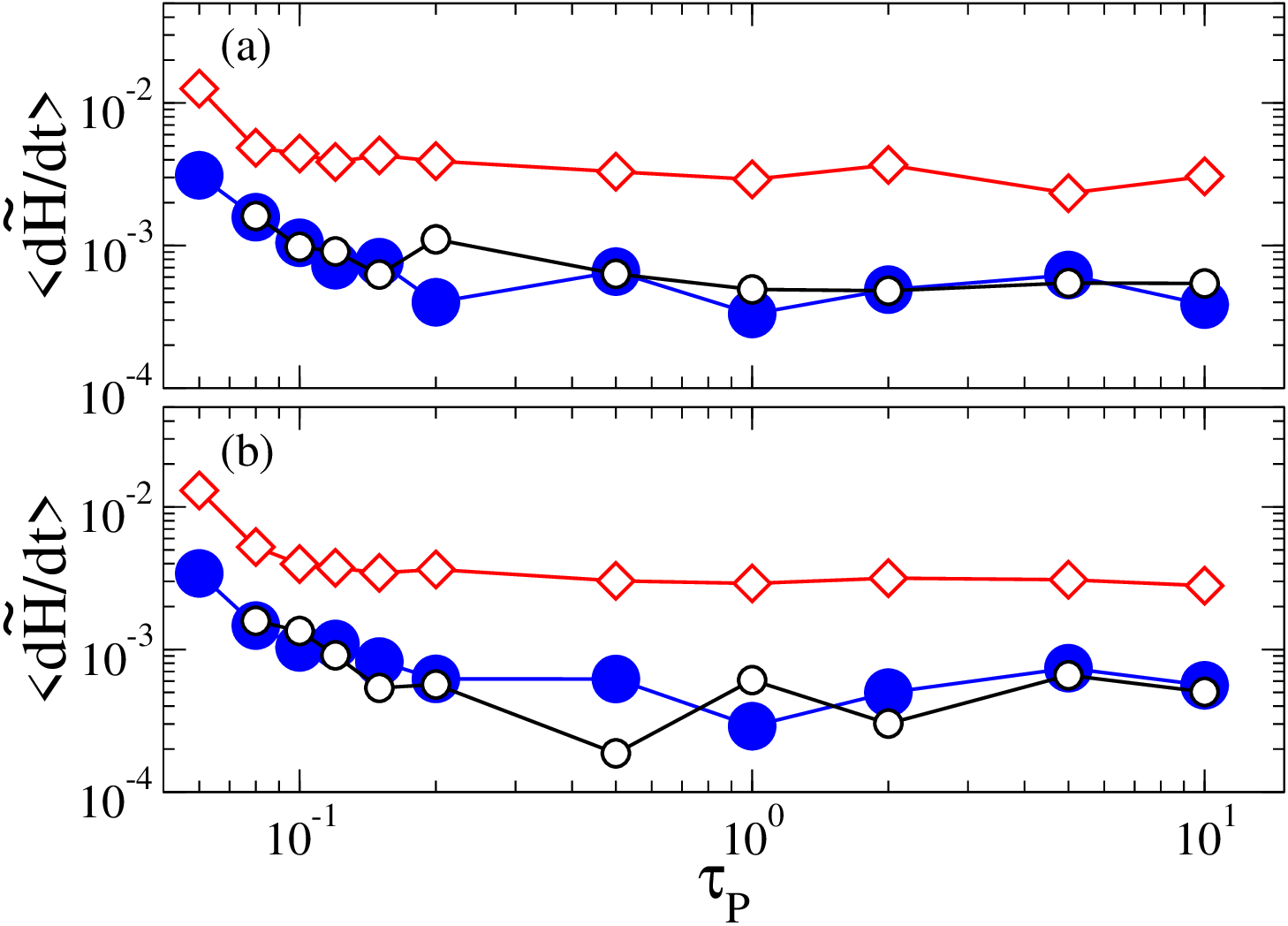}
\caption{Drift of the effective enthalpy $\langle\frac{d\tilde{H}}{dt}\rangle$
plotted as a function of the barostat relaxation time $\tau_P$
for the Lennard-Jones liquid (a) and solid (b) systems at the triple point,
for stochastic velocity rescaling (filled circles),
Nos\'{e}-Hoover (open circles) and Langevin (diamonds) algorithms.
The thermostat relaxation time is $\tau_T=0.2$ and the timestep is
$\Delta t=0.005$.
All quantities are in Lennard-Jones reduced units.}
\label{flux_tp}
\end{figure}
We now consider the choice of $\tau_P$.
We recall that there is a relationship between the mass of the
particles (and of the barostat) and the time-step. As an example, a
four-fold decrease of all the masses produces exactly the same trajectory
as a two-fold increase of the time-step, thus giving a two times faster exploration.
In the same way, for a fixed value of the particles mass,
a lower barostat mass will provide a faster volume evolution at the price of a lower accuracy.
Thus, the choice of the barostat mass, $W=N_f\beta^{-1}\tau_P^2$, is
based on a compromise between a small mass and a reasonable effective-enthalpy drift,
and is very similar to the choice of $\Delta t$.
In Fig.~\ref{flux_tp} we show the dependence of the effective-enthalpy drift on $\tau_P$.
Since the drift induced by a too small $\tau_P$ is due
to sampling errors of a single degree of freedom (the volume),
it is partially hidden by the larger
drift due to the inexact integration of the particles degrees of freedom.
However, it is clear that for $\tau_P\ge 0.5$ the drift due
to the volume is negligible.
We notice that in none of the presented simulations we have observed
an appreciable decoupling between the barostat and the internal degrees of
freedom in the case of small $\tau_P$.
We also computed the autocorrelation time of a few relevant quantities,
and we noticed that the sampling efficiency always increases as $\tau_P$
is decreased. Thus, the final rule turns out to be a $\tau_P$ which is
as small as possible, but with a reasonable effective-enthalpy conservation.
We thus choose $\tau_P=0.5$ to continue our tests.

\subsection{Thermostat relaxation time and statistical efficiency}

\begin{figure}[h]
\includegraphics[width=0.48\textwidth,clip]{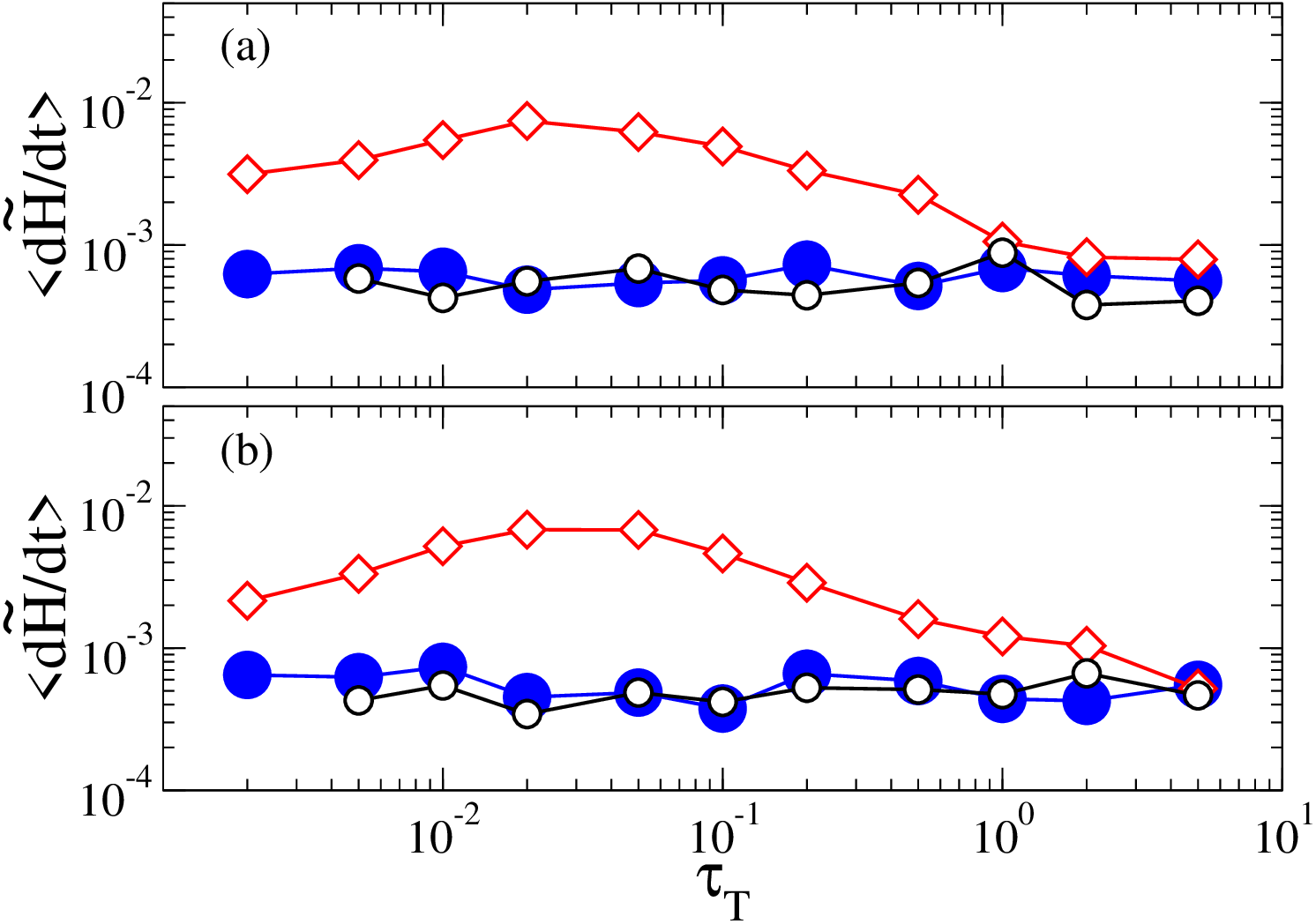}
\caption{Drift of the effective enthalpy $\langle\frac{d\tilde{H}}{dt}\rangle$
plotted as a function of the thermostat relaxation time
$\tau_T$ for the Lennard-Jones liquid (a) and solid (b) systems at the triple point,
for stochastic velocity rescaling (filled circles),
Nos\'{e}-Hoover (open circles) and Langevin (diamonds) algorithms.
The timestep is $\Delta t=0.005$ and the barostat relaxation time is $\tau_P=0.5$.
All quantities are in Lennard-Jones reduced units.}
\label{flux_tt}
\end{figure}
We now proceed to a systematic study of the effects of the other relevant
simulation parameter, namely the thermostat relaxation time $\tau_T$.
To this aim, we fix a timestep $\Delta t=0.005$ and a barostat relaxation time
$\tau_P=0.5$, and perform simulations for a wide range of $\tau_T$.
As it is seen in Fig.~\ref{flux_tt}, the effect of $\tau_T$
on the effective-enthalpy drift is rather weak, at least
for the two global schemes. Indeed, for these
choices of the parameters, the origin of the drift
is not in the thermostat but in the constant-enthalpy step.
The drift in Langevin is larger than that of the other
schemes.
For $\tau_T$ we cannot use the same criterion that we used to optimize
$\tau_P$. Indeed the two stochastic schemes (rescaling
and Langevin) are stable for any possible choice of $\tau_T$. This is 
due to the fact that, in these two cases, the thermostat equations are
linear and can be integrated in an analytic fashion
(see Appendix).
This is not true for the NH scheme,  which turns out to be
be unstable for $\tau_T\approx 0.02$, at least using the present
integration scheme.
Moreover, it is not obvious that a smaller
$\tau_T$ should lead to faster sampling, as it will be seen in the following.

We thus proceed in an analysis of the statistical efficiency as a function of the thermostat
parameter,
similar to what has been done for the $NVT$ ensemble.\cite{buss-parr08cpc}
A quantitative measure of the time needed for performing ensemble averages of
a given quantity $X$ is given by its autocorrelation time $\tau_X$
(see e.g.~Ref.~\onlinecite{land-bind2005book}).
Since a common goal of $NPT$ simulation is the calculation of averages
of quantities like the enthalpy and the volume of the system, we consider here
the cases where $X=H$ and $X=V$.
The autocorrelation time of the total enthalpy also provides a rough estimate
of the time needed for the equilibration procedure.
However, it must be noticed that to achieve a fast equilibration also
the autocorrelation time of the \emph{fluctuations} of the enthalpy should
be short.
Thus, we also calculate the correlation time for the
fluctuations of these quantities, obtained with $X=\Delta H^2\equiv(H-\langle H\rangle)^2$
and $X=\Delta V^2\equiv(V-\langle V\rangle)^2$ respectively. 
The first one is related to the bulk modulus and
the second to the heat capacity at constant pressure.
Thus, these autocorrelation times provide the statistical efficiency in the
calculation of these important physical observables.

\begin{figure}[h]
\includegraphics[clip,width=0.48\textwidth]{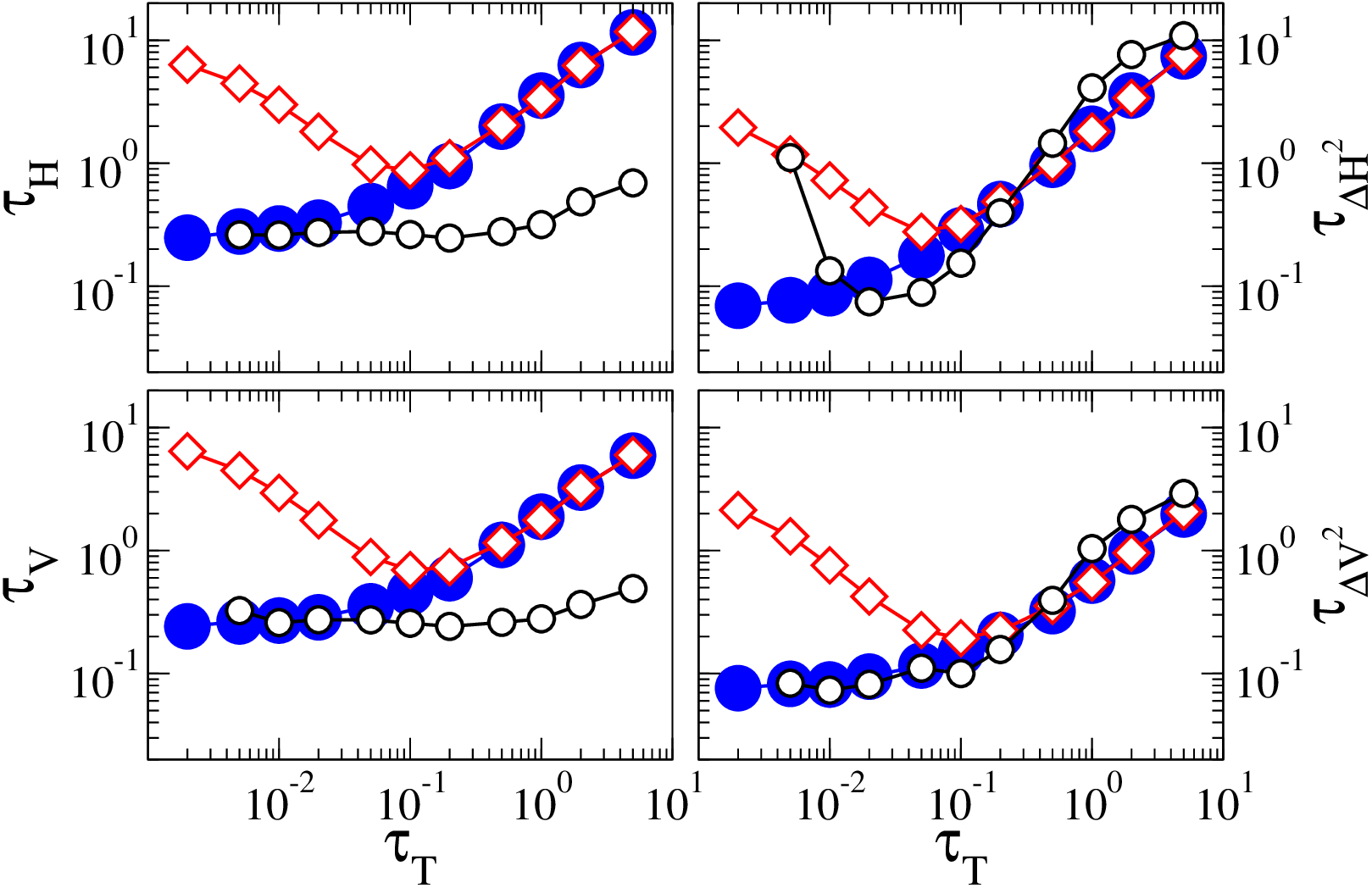}
\caption{Autocorrelation times of the average enthalpy $\tau_{H}$,
of the enthalpy fluctuations $\tau_{\Delta H^2}$,
of the volume $\tau_{V}$ and of the volume fluctuations $\tau_{\Delta V^2}$,
plotted against thermostat relaxation time $\tau_T$,
for the Lennard-Jones liquid at the triple point,
for stochastic velocity rescaling (filled circles),
Nos\'{e}-Hoover (open circles) and Langevin (diamonds) algorithms.
The timestep is $\Delta t=0.005$ and the barostat relaxation time is $\tau_P=0.5$.
All quantities are in Lennard-Jones reduced units.
\label{fig-auto-liquid}
}
\end{figure}
\begin{figure}[h]
\includegraphics[clip,width=0.48\textwidth]{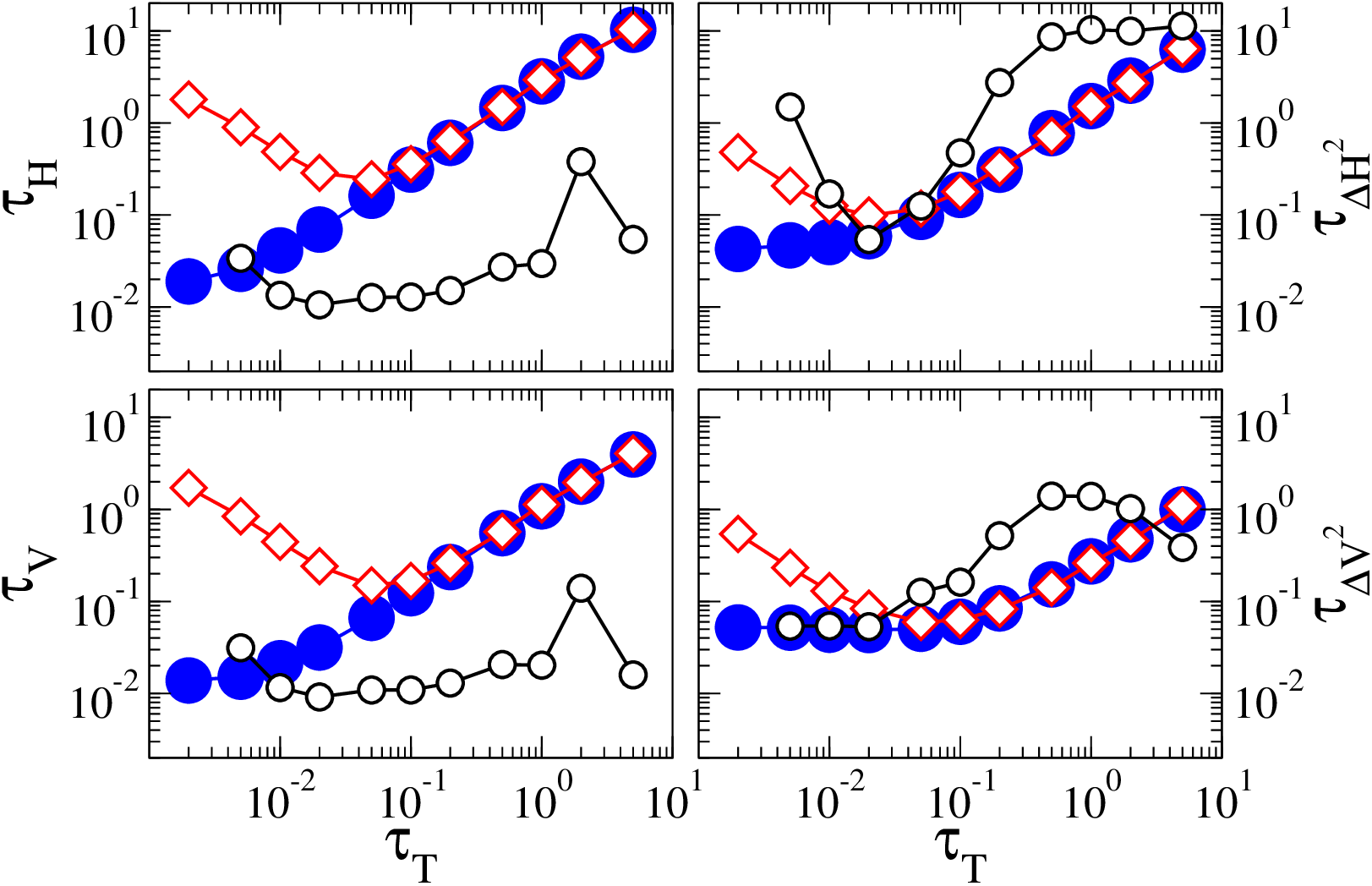}
\caption{Autocorrelation times of the average enthalpy $\tau_{H}$,
of the enthalpy fluctuations $\tau_{\Delta H^2}$,
of the volume $\tau_{V}$ and of the volume fluctuations $\tau_{\Delta V^2}$,
plotted against thermostat relaxation time $\tau_T$,
for the Lennard-Jones solid at the triple point,
for stochastic velocity rescaling (filled circles),
Nos\'{e}-Hoover (open circles) and Langevin (diamonds) algorithms.
The timestep is $\Delta t=0.005$ and the barostat relaxation time is $\tau_P=0.5$.
All quantities are in Lennard-Jones reduced units.
\label{fig-auto-solid}
}
\end{figure}
The simulation results for a number of runs are summarized in 
Fig.~\ref{fig-auto-liquid} and Fig.~\ref{fig-auto-solid} for the liquid and the solid respectively.
In all cases, the performance of stochastic rescaling is equal or better than that of
the local Langevin thermostat.
For large $\tau_T$, the two stochastic schemes have the same behavior. For small
$\tau_T$, the performance of the local Langevin scheme is worse.
This happens for both the solid and liquid systems, and indicates
that tuning the performance of Langevin thermostat is very difficult since
it is efficient only for a limited friction range.
These problems in Langevin thermostat are related to the fact that it is local,
thus it disturbs considerably the trajectory, hindering
diffusion (in the liquid) or thermalization
of the slow modes (in the solid).
This is in agreement to what has been observed
in Ref.~\onlinecite{buss-parr08cpc} for a liquid in the $NVT$ ensemble.

On the other hand, the Nos\'{e}-Hoover thermostat is global and, similarly to the
velocity rescaling, has a very small impact on the trajectory.
In the present calculations, performed at the triple point,
we don't expect the solid to be harmonic enough to introduce
ergodicity problems.
Its performances are excellent when looking at the autocorrelation time
of enthalpy ($\tau_H$) and volume ($\tau_V$), for both the solid and the liquid.
For the solid, a resonance effect leads to a peak in these autocorrelation times
for $\tau_T\approx 2$.
However, fluctuations of the enthalpy ($\tau_{\Delta H^2}$) are well thermalized
only for a limited range of choices for $\tau_T$. This is related to the fact
that the thermostat is optimally coupled only to the modes with a given frequency,
and is not able to thermalize the others.
The fact that the average enthalpy has a short autocorrelation time is due
to cancellations in the autocorrelation function, and is a signature
of the enthalpy oscillations around the correct value.
The damping of these oscillations can be rather slow if $\tau_T$
is not chosen properly.

The stochastic rescaling scheme combines the stability of the
stochastic Langevin with the efficiency of the global
Nos\'{e}-Hoover.
The coupling parameter $\tau_T$ in the stochastic rescaling scheme has the meaning of
thermostat-strength, and a smaller $\tau_T$ always guarantees a better efficiency.
On the contrary, in the Nos\'{e}-Hoover scheme $\tau_T$ also determines
which modes are optimally coupled with the thermostat, and it is more
difficult to tune.
Finally, the stochastic rescaling
is comparable with the Nos\'{e}-Hoover in computation of averages for small $\tau_T$,
surpassing it in computation of fluctuations.

\subsection{Dynamical properties}

It is also interesting to check if the rescaling scheme preserves the dynamics.
To investigate this point, we consider the calculation of dynamical properties.

\begin{figure}[h]
\includegraphics[width=0.5\textwidth,clip]{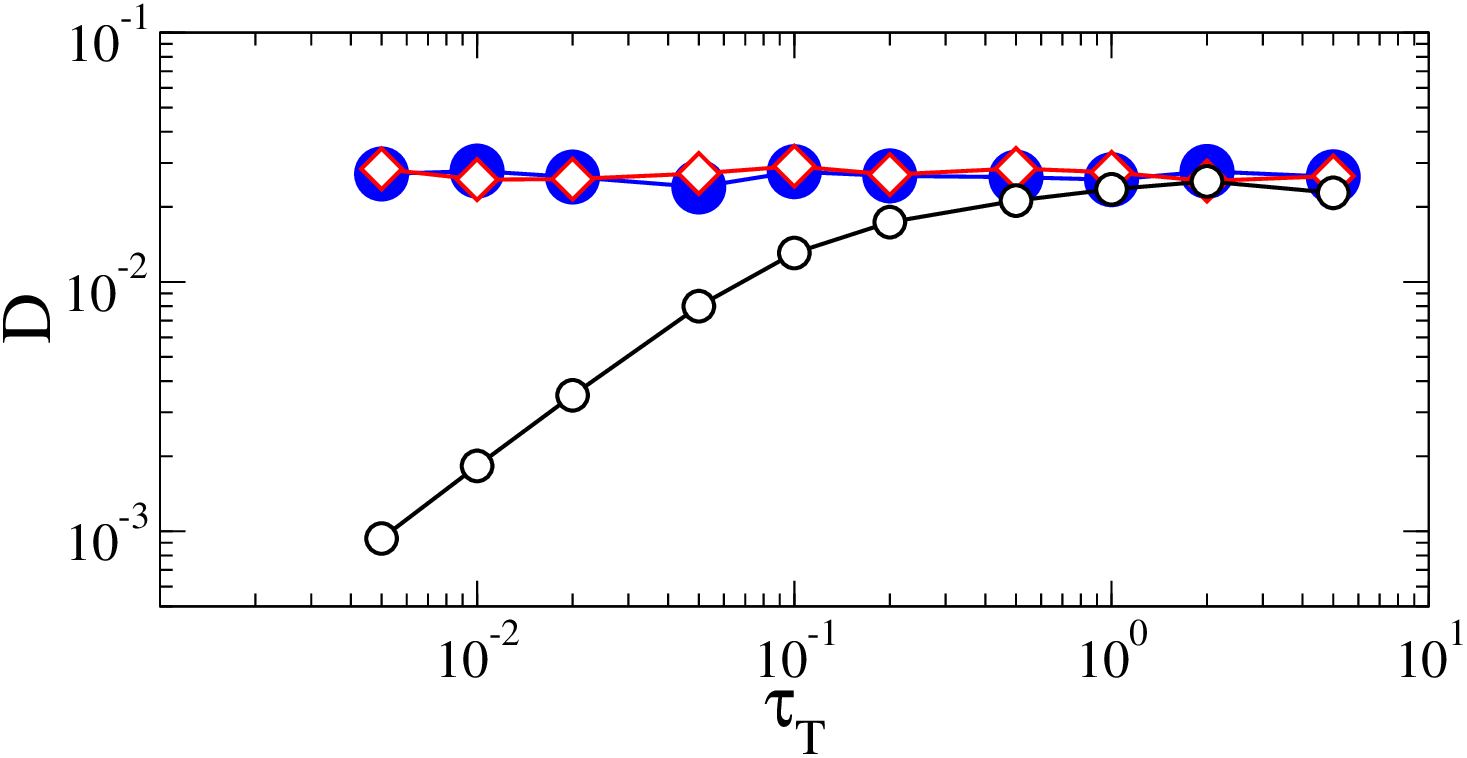}
\caption{Diffusion coefficient
plotted against thermostat relaxation time $\tau_T$,
for the Lennard-Jones liquid at the triple point,
for stochastic velocity rescaling (filled circles),
Nos\'{e}-Hoover (open circles) and Langevin (diamonds) algorithms.
The diffusion coefficient computed from the corresponding $NVE$ run is 0.03.
The timestep is $\Delta t=0.005$ and the barostat relaxation time is $\tau_P=0.5$.
All quantities are in Lennard-Jones reduced units.
\label{diffusion}
}
\end{figure}
For the LJ liquid, we calculate the self-diffusion coefficient $D$ 
from the mean square displacement.
In order to overcome the practical problem of computing the mean square displacement
in a variable cell, first we calculate it using the scaled coordinates
$V^{-1/3}\bm{q}_i$, then we scale the obtained diffusion coefficient
of a factor $\langle V\rangle^{2/3}$ to recover the correct value.
The diffusion coefficient as a function of the thermostat relaxation time $\tau_T$ is shown on Fig.~\ref{diffusion}.
It is evident that both global schemes (velocity rescaling and Nose-Hoover)
do not affect the diffusion. In particular, the calculated value ($D=0.03$) is in agreement
with that obtained in microcanonical simulations in similar conditions.
On the other hand, when the local Langevin scheme is used with a too high friction
the diffusion is hindered.
Remarkably, the choices of $\tau_T$ which lead to a slower diffusion correspond
to the choices that lead to a worse sampling (see the previous Subsection).
These time can also be compared with the typical collision time, defined as
the velocity autocorrelation time $\tau_v$. It is easy to show that
$\tau_v$ is related to $D$ by $\tau_v=\beta m D$.
In the limit of large $\tau_T$ (small friction), it turns out
to be $\tau_v=0.04$. When the collisions with the thermal bath have the
same frequency that the natural interparticle collisions, the thermostat
turns out to be the bottleneck for diffusion.

\begin{figure}[h]
\includegraphics[width=0.5\textwidth,clip]{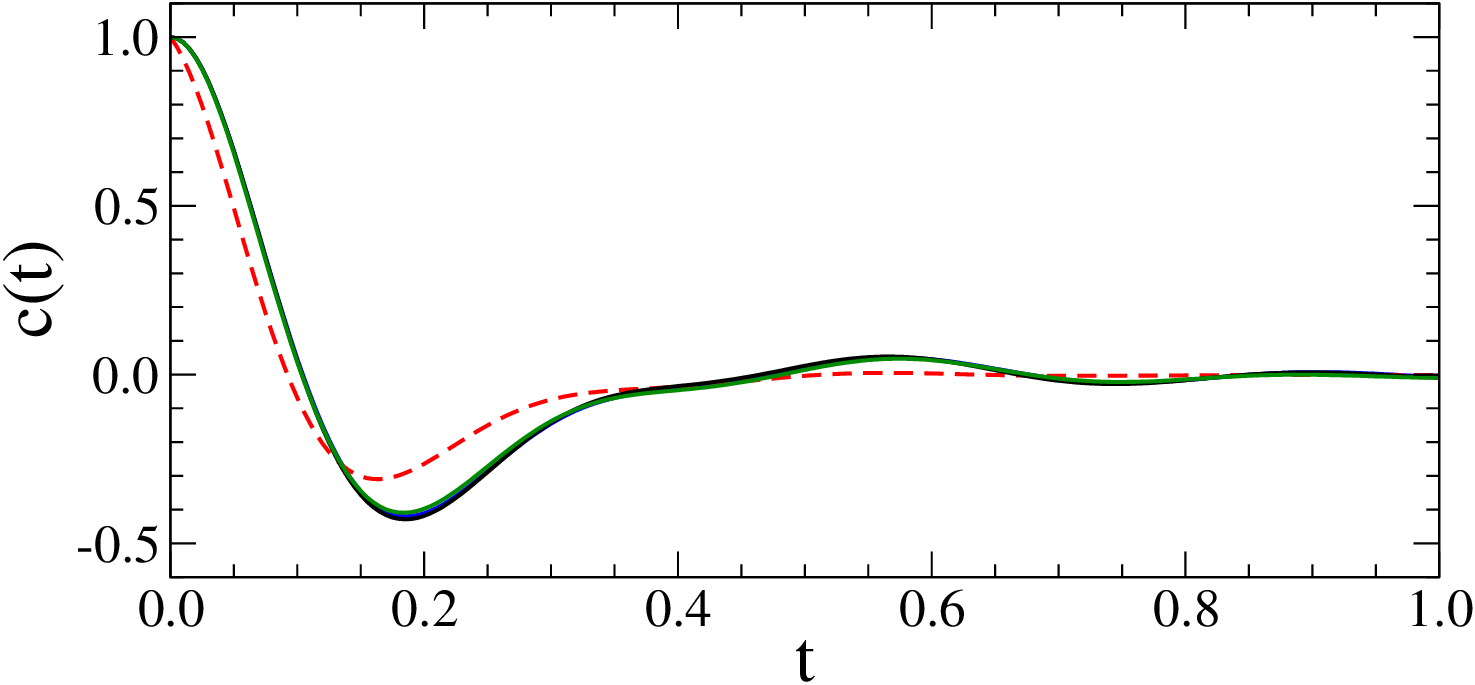}
\caption{Normalized velocity-velocity autocorrelation function $c(t)$ for the Lennard-Jones
solid at the triple point. The three overlapping solid lines, which are 
almost indistinguishable, are obtained using
stochastic velocity rescaling, Nos\'{e}-Hoover
and microcanonical dynamics. The dashed line is obtained using the Langevin thermostat.
All the thermostats are used with a thermostat relaxation time $\tau_T=0.1$
(for Langevin $\gamma=5$).
The timestep is $\Delta t=0.005$ and the barostat relaxation time is $\tau_P=0.5$.
All quantities are in Lennard-Jones reduced units.
\label{velocity}
}
\end{figure}
In the solid, where the diffusion coefficient is vanishing,
we opt for the calculation of the full velocity-velocity autocorrelation function.
The results are shown in Fig.~\ref{velocity}, where the autocorrelation function is plotted for
all the considered thermostats, using a reasonable thermostat relaxation time $\tau_T=0.1$,
and for a reference microcanonical simulation at similar conditions.
From the plot it is clear that the global schemes do not affect significantly 
the dynamical properties. Even if these simulations are performed in
the $NPT$ ensemble, they can be used to calculate correct dynamical
properties.
This is not true for the local Langevin scheme, where the autocorrelation function is strongly
affected by the thermostat.
This is an additional
confirmation that the impact on the dynamical properties is not due to the stochastic nature
but to the locality of the Langevin thermostat.

\section{Conclusions}
A new method has been introduced to perform MD simulations at
constant temperature and pressure.
The method is stochastic but does not affect the dynamical properties.
The concept of effective energy, which provides a constant of motion
for stochastic algorithms and which was already introduced in
previous papers,\cite{buss+07jcp,buss-parr07pre} has been extended
in the present scheme to an effective enthalpy.
A systematic analysis of the role played by the control parameter has been shown,
focused on the choice of the thermostat and barostat relaxation times.
Moreover, a comparison of the new scheme with standard
Nos\'{e}-Hover and Langevin barostats has been shown.
The new scheme appear to be simpler to use and easier to control.
Further work is required to apply the presented algorithm to
the case of nonequilibrium MD, where the Gibbs' ensemble
does not hold.\cite{evan+83pra}

\appendix
\section{Integration scheme}
At variance with Ref.~\onlinecite{buss+07jcp}, we follow here a time-reversible
scheme
which is consistent with the Trotter decomposition
scheme.\cite{trot59pams,tuck+92jcp}
The Trotter decomposition is usually adopted
to decompose the Liouville (or Fokker-Planck) operator
which describes the dynamics of an ensemble of realizations.
For instance, if the exact solution of the Liouville equation
for a finite time increment $\Delta t$ is
$\rho(t+\Delta t)=e^{-(\hat{L}_1+\hat{L}_2+\hat{L}_3)\Delta t}\rho(t)$,
its approximated solution is
$\rho(t+\Delta t)\approx e^{-\hat{L}_1\frac{\Delta t}{2}} e^{-\hat{L}_2\frac{\Delta t}{2}} e^{-\hat{L}_3\Delta t}
e^{-\hat{L}_2\frac{\Delta t}{2}} e^{-\hat{L}_1\frac{\Delta t}{2}}\rho(t)$.
Equivalently, one can state that if the equations of motion are in the
form $\dot{x}=G_1(x)+G_2(x)+G_3(x)$ an approximate
propagator can be obtained applying in the proper order
the analytical solutions of the equations
$\dot{x}=G_1(x)$, $\dot{x}=G_2(x)$ and $\dot{x}=G_3(x)$.
A necessary condition is that the three decoupled
equations can be solved exactly.

We here decompose the equations of motion as a sum of three
parts: the first is the propagation of the thermostat,
and the others are two equations of motion whose
sum is equivalent to the $NPH$ dynamics.
The flow of the integration scheme is as follows: 
\begin{enumerate}
\item
Propagate the thermostat for a time $\Delta t/2$.
\item Propagate velocities for a time $\Delta t/2$, according to
\begin{subequations}
\label{eq-trotter-v}
\begin{align}
\dot{\bm r}_i=&0
\\
\dot{\bm \pi}_i=&{\bm f}_i
\\
\dot{V}=&0
\\
\dot{\eta}=&\frac{3[V(P_{int}-P_{ext})+2\beta^{-1}]}{W}
\end{align}
\end{subequations}
\item Propagate positions and velocities for a time $\Delta t$, according to
\begin{subequations}
\label{eq-trotter-r}
\begin{align}
\dot{\bm r}_i=&\frac{\bm \pi_i}{m_i}+\eta {\bm r}_i
\\
\dot{\bm \pi}_i=&-\eta {\bm \pi_i}
\\
\dot{V}=&3V\eta
\\
\dot{\eta}=&0
\end{align}
\end{subequations}
\item Propagate again velocities for a time $\Delta t/2$, as in step 2.
\item Propagate again the thermostat for a time $\Delta t/2$, as in step 1.
\end{enumerate}
The sum of Eqs.~\eqref{eq-trotter-v} and~\eqref{eq-trotter-r} is 
equivalent to the original equations of motion~\eqref{eq-nph-motion}.
This choice for the splitting is arbitrary, but it is motivated by
the fact that Eqs.~\eqref{eq-trotter-v} and~\eqref{eq-trotter-r} can be solved
analytically also for finite time.

We now give explicit expressions for the propagation of the single stages.
Stages 2 and 4 are propagated solving Eqs.~\eqref{eq-trotter-v}:
\begin{subequations}
\label{eq-propagation-p}
\begin{multline}
\eta(t+\frac{\Delta t}{2})= \eta(t) +
\frac{3[V(P_{int}(t)-P_{ext})+2\beta^{-1}]}{W}
\frac{\Delta t}{2} \\
          + \sum_i\frac{{\bm f}_i(t) \cdot {\bm p}_i(t)}{Wm_i}
                               \left(\frac{\Delta t}{2}\right)^2
          + \sum_i\frac{|{\bm f}_i(t)|^2}{Wm_i}
               \frac{1}{3}     \left(\frac{\Delta t}{2}\right)^3
\end{multline}
\begin{equation}
{\bm p}_i(t+\frac{\Delta t}{2}) = {\bm p}_i(t) + {\bm f}_i(t) \frac{\Delta t}{2}.
\end{equation}
\end{subequations}
The instantaneous pressure here is calculated as
\begin{multline}
P_{int}(t)=
\frac{2K}{3V}
- \frac{d U}{d V}= \\
\frac{1}{3V}\left[
\sum_i\frac{|{\bm p}_i(t)|^2}{m_i}+
\sum_{ij}{\bm f}_{ij}(t)\cdot{\bm r}_{ij}(t)
\right]-\frac{\partial U}{\partial V}
\end{multline}
where ${\bm f}_{ij}(t)$ is the force between particles $i$ and $j$
and ${\bm r}_{ij}$ is their distance, evaluated by taking into account the periodic
boundary conditions within the
minimum-image convention.\cite{alle-tild87book,louw-baer06cpl}
The last term $\partial U/\partial V$ takes into
account any explicit dependence of the potential energy on the volume,
such as the long range corrections.\cite{alle-tild87book,mart+94jcp}

Stage 3 is propagated solving Eqs.~\eqref{eq-trotter-r}:
\begin{subequations}
\label{eq-propagation-q}
\begin{equation}
{\bm r}_i(t+\Delta t) = \frac{\sinh[\eta(t) \Delta t]}{\eta(t)} \frac{{\bm p}_i(t)}{m_i}
+e^{\eta(t)\Delta t} {\bm r}_i(t)
\end{equation}
\begin{equation}
V(t+\Delta t) = e^{3\eta(t)\Delta t} V(t)
\end{equation}
\begin{equation}
{\bm p}_i(t+\Delta t) = e^{-\eta(t)\Delta t} {\bm p}_i(t).
\end{equation}
\end{subequations}

It is also useful to evaluate the compressibility associated to this propagators,
defined as the Jacobian of the corresponding transforms.
The transforms in Eqs.~\eqref{eq-propagation-p} and~\eqref{eq-propagation-q} are
linear, with Jacobian equal to respectively 1 and $e^{3\eta(t)\Delta t}=V(t+\Delta t)/V(t)$.
Thus, when a volume element in the phase space is transformed by Eqs.~\eqref{eq-propagation-p}
and~\eqref{eq-propagation-q} its measure is changed by a a factor $V(t+\Delta t)/V(t)$.
This contribution is crucial for a proper evaluation
of detailed-balance violations as it is done in Subsection~\ref{sec-sampling-errors}.

We now describe the propagation of the thermostat stages (1 and 5).
The thermostat can be either velocity rescaling, Langevin or Nos{\'e}-Hoover.
We here simply state explicitly the integration schemes for the stochastic ones,
without including any derivation.
The complete derivations can be found in the original papers.

In the case of velocity rescaling,
the propagation of the thermostat reads:\cite{buss+07jcp,buss-parr08cpc}
\begin{subequations}
\begin{equation}
{\bm p}_i(t+\Delta t/2)
=
\alpha(t) {\bm p}_i(t),
\end{equation}
\begin{equation}
\eta(t+\Delta t/2)
=
\alpha(t) \eta(t)
\end{equation}
where
\begin{multline}
\alpha^2(t) = 
c+\frac{(1-c)[S_{N^*_f-1}(t)+R^2(t)]\bar{K}^*}{N^*_fK^*(t)} \\
+ 2R(t)\sqrt{\frac{c(1-c)\bar{K}^*}{N^*_fK^*(t)}}.
\end{multline}
and
\begin{equation}
\label{eq-alpha-sign}                                                               
\text{sign} [\alpha(t)]=\text{sign}\left[R(t)+\sqrt{\frac{cN^*_fK^*(t)}{(1-c)\bar{K}^*}}\right].
\end{equation}
\end{subequations}
Here $c=e^{-\Delta t /(2\tau_T)}$, $R(t)$ is a Gaussian number
with unitary variance and $S_{N^*_f-1}$ is the sum of $N^*_f-1$
independent, squared, Gaussian numbers.
$N^*_f=3N-3+1$ does not include the center of mass but includes the volume degree of freedom.

In the case of Langevin dynamics, the propagation of the thermostat reads:\cite{buss-parr07pre,adja+06epjb}
\begin{subequations}
\begin{equation}
{\bm p}_i(t+\Delta t/2)
=
c
{\bm p}_i(t)+\sqrt{\frac{m_i(1-c^2)}{\beta}}{\bm R}_i(t)
\end{equation}
\begin{equation}
\eta(t+\Delta t/2)=
c\eta(t)+\sqrt{\frac{(1-c^2)}{\beta W}}R(t)
\end{equation}
\end{subequations}
where $c=e^{-\gamma\Delta t/2}$. Here the ${\bm R}_i(t)$ are $N$ independent
vectors of normalized Gaussian numbers.
After the application of the thermostat we remove the center-of-mass velocity,
so as to have an ensemble exactly equivalent to that obtained with
velocity rescaling.

\end{document}